\documentclass[twocolumn,showpacs,preprintnumbers,amsmath,amssymb]{revtex4}
\usepackage{mathrsfs}
\usepackage{bbm}

\usepackage{graphicx}
\usepackage{dcolumn}
\usepackage{bm}

\begin{document}
\title{First-principles study of ground state properties and high pressure behavior of ThO$_{2}$}
\author{Baotian Wang$^{1,2}$}
\author{Hongliang Shi$^{2,3}$}
\author{Weidong Li$^{1}$}
\author{Ping Zhang$^{2}$}
\email{zhang_ping@iapcm.ac.cn} \affiliation{$^{1}$Institute of
Theoretical Physics and Department of Physics, Shanxi University,
Taiyuan 030006, People's Republic of China\\$^{2}$LCP, Institute of
Applied Physics and Computational Mathematics, Beijing 100088,
People's Republic of China\\$^{3}$SKLSM,Institute of Semiconductors,
Chinese Academy of Sciences, People's Republic of China}

\pacs{61.50.Ah, 61.50.Ks, 71.15.Mb, 63.20.dk}

\begin{abstract}
The mechanical properties, electronic structure and phonon
dispersion of ground state ThO$_{2}$ as well as the structure
behavior up to 240 GPa are studied by using first-principles
density-functional theory. Our calculated elastic constants indicate
that both the ground state fluorite structure and high pressure
cotunnite structure of ThO$_{2}$ are mechanically stable. The bulk
modulus, shear modulus, and Young's modulus of cotunnite ThO$_{2}$
are all smaller by approximately 25\% compared with those of
fluorite ThO$_{2}$. The Poisson's ratios of both structures are
approximately equal to 0.3 and the hardness of fluorite ThO$_{2}$ is
27.33 GPa. The electronic structure and bonding nature of fluorite
ThO$_{2}$ are fully analyzed, which show that
the Th-O bond displays a mixed ionic/covalent character. The valence
of Th and O ions in fluorite ThO$_{2}$ can be represented as
Th$^{3.834+}$ and O$^{0.452-}$. The phase transition from the fluorite to cotunnite structure is calculated to be at the pressure of
26.5 GPa, consistent with recent experimental measurement by Idiri \emph{et al}. \cite{Idiri}. For the cotunnite phase it is further predicted
that an isostructural transition takes place in the pressure region of 80 to 130 GPa.
\end{abstract}
\maketitle

\section{INTRODUCTION}
Besides uranium and plutonium, thorium is also one kind of important
nuclear materials. Metal thorium and its compounds have been widely
investigated both experimentally and theoretically since metal
thorium was found in 1828. Among thorium compounds, thorium dioxide
ThO$_{2}$, which is a stable diamagnetic transparent insulating
solid, has attracted special attention. In addition to its usage as
an important nuclear fuel material, thorium dioxide has also been
used as a solid-state electrolyte. In particular, due to its
prominent hardness, ThO$_{2}$ has potential interests as an optical
component material and laser host.

Recently, there has occurred in the literature a series of
experimental reports on pressure-induced phase transition of
ThO$_{2}$ \cite{Jaya,Dancausse,Idiri}. At ambient pressure,
ThO$_{2}$ crystallizes in the (CaF$_{2}$-type) fluorite structure
with space group \emph{Fm$\bar{3}$m} (No. 225). By using the energy
dispersive x-ray diffraction (EDXRD) method, Dancausse \emph{et al}.
\cite{Dancausse} reported that at 40 GPa, ThO$_{2}$ undergo a phase
transition to an orthorhombic structure of cotunnite (PbCl$_{2}$)
type with space group \emph{Pnma} (No. 62). Later, through improving
experimental measurement technique, Idiri \emph{et al}. \cite{Idiri}
observed that this phase transition really begins around 30 GPa, and
the two phases coexist in a wide pressure range of nearly 20 GPa.
Additional high-pressure Raman spectroscopy measured by Jayaraman
\emph{et al}. \cite{Jaya} also suggested that ThO$_{2}$ starts to
transform into the cotunnite structure around 30 GPa.

In contrast to the above-mentioned extensive experimental
investigation, a systematic theoretical study on the phase
transition of ThO$_{2}$ at high pressure is still lacking. In
particular, considering the obvious discrepancy between two
experimental groups \cite{Dancausse,Idiri} involving the transition
pressure, such a theoretical investigation from basis quantum
mechanics is not only complementary but also indispensable.
Motivated by this observation, in this paper, we present a
first-principles study by calculating the total energies and
enthalpies of ThO$_{2}$ at its experimentally established
crystalline phases. Our calculation shows that the transition
pressure is around 30 GPa, consistent with the recent experiment by
Idiri \emph{et al}. \cite{Idiri}.

The other task for this paper is to theoretically present a thorough
description of physical, mechanical and chemical bonding properties
of ThO$_{2}$ at its ground state of the fluorite phase. To date,
theoretical studies of the ground-state behavior of ThO$_{2}$ are
very scarce in literature \cite{Kelly,Harding,Staun,Li,Kanchana} and
some of them are even inconsistent with the experimental data to a
large extent. For example, without taking into account the $5f$
state, Kelly \emph{et al}. \cite{Kelly} calculated the bulk modulus
$B$ of ThO$_{2}$ to be 290 GPa, which is far from the experimental
value of 198 GPa. On the other hand, based on a purely ionic model,
Harding \emph{et al}. \cite{Harding} obtained $B$=175 GPa, which,
compared to the experimental data, clearly underestimate the binding
interaction in the material to a large extent. This as whole
encourages us to theoretically report a systematic investigation on
the ground-state properties of ThO$_{2}$ at its ambient phase.
Besides the consistency with the existing experimental data, we
expect that the new results for the first time predicted in this
paper and experimentally inaccessible at present, such as the
hardness, the phonon spectrum, and the charge transfer of the Th-O
bond, will be of great help for a primary understanding of
ThO$_{2}$. The rest of this paper is arranged as follows. In Sec. II
the computational method is briefly described. In Sec. III and Sec.
IV we present our calculated results, some of which are made
comparison with available experimental and theoretical results. In
Sec. V we summarize the conclusions of this work.

\section{computational method}
Our total energy calculations are carried out by employing the
plane-wave basis pseudopotential method as implemented in Vienna
\textit{ab initio} simulation package (VASP) \cite{Kresse3}. The
exchange and correlation effects are described by the DFT within the
generalized gradient approximation (GGA) \cite{GGA}. The electron
and core interactions are included using the frozen-core projected
augmented wave (PAW) approach which combines the accuracy of
augmented-plane-wave methods with the efficiency of the
pseudo-potential approach \cite{PAW}. The thorium
6s$^{2}$7s$^{2}$6p$^{6}$6d$^{1}$5f$^{1}$ and the oxygen
2s$^{2}$2p$^{4}$ electrons are treated as valence electrons. Note
that although the $5f$ state are empty in elemental Th, this level
turns to evolve into a hybridization with O $2p$ state in the
valence band, as well as to prominently contributes to the
conduction band (see Fig. 3 below). We use a 9$\times$9$\times$9
\emph{k}-point grid with Monkhorst-Pack scheme \cite{Monk} for
fluorite ThO$_{2}$ and 9$\times$15$\times$9 \emph{k}-point grid for
cotunnite structure. Electron wave function is expanded in plane
waves up to a cutoff energy of 500 eV, and all atoms are fully
relaxed until the quantum mechanical forces become less than 0.02
eV/\AA.

To obtain optimized lattice constants of ground state ThO$_{2}$,
first, we calculate several total energies at different lattice
constants around the experimental value. Then we fit the
energy-volume data to the third-order Birch-Murnaghan equation of
state (EOS) \cite{Birch} to give the theoretical equilibrium volume,
minimum energy, bulk modulus \emph{B}, and pressure derivative of
the bulk modulus \emph{B$'$}. On the other hand, the bulk modulus
\emph{B}, shear modulus \emph{G}, Young's modulus \emph{E},
Poisson's ratio $\upsilon$, can also be derived from the elastic
constants. We find that the bulk modulus \emph{B} obtained by these
these two ways are in good agreement, indicating that our
calculations are self-consistent.

\section{ground state properties for fluorite-type thorium dioxide}

\subsection{Atomic structure and mechanical properties}
\begin{figure}[tbp]
\begin{center}
\includegraphics[width=1.0\linewidth]{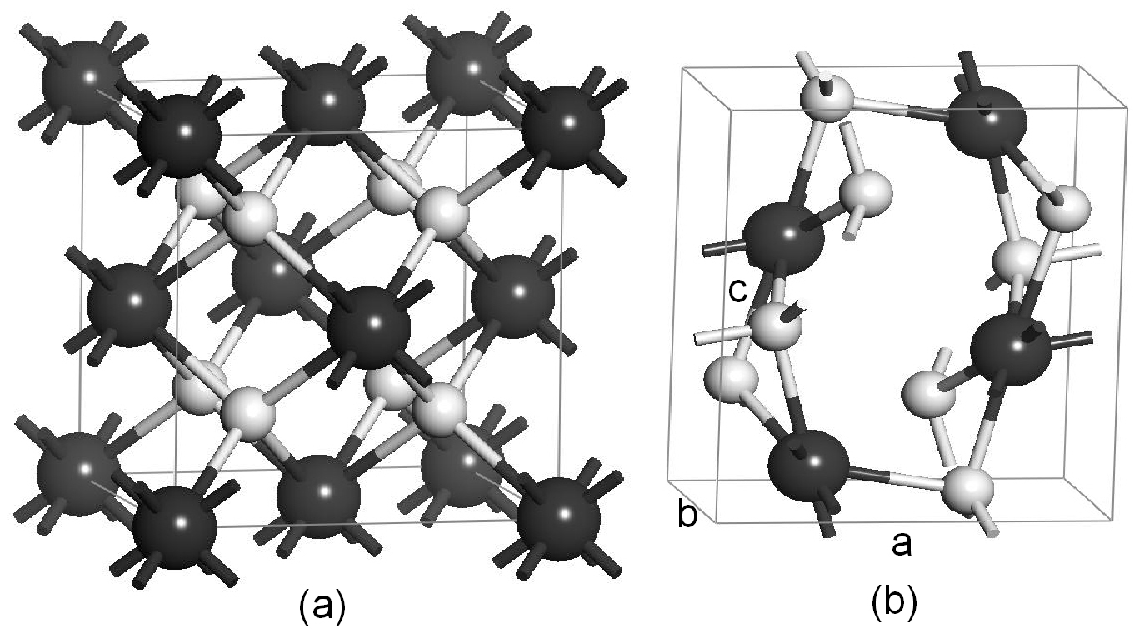}
\end{center}
\caption{Cubic unit cell for ThO$_{2}$ in space group
\emph{Fm$\bar{3}$m} (a) and orthorhombic unit cell in space group
\emph{Pnma} (b), larger black spheres stand for Th atoms and the
smaller white O.} \label{fig1} \label{structure}
\end{figure}%

\begin{table*}
\caption{ Calculated lattice parameter, elastic constants, bulk
modulus \emph{B}, pressure derivative of the
bulk modulus \emph{B$^{'}$}, shear modulus \emph{G}, Young's modulus
\emph{E}, Poisson's ratio $\upsilon$, hardness, and energy band gap
(\emph{E$_{g}$}) for cubic ThO$_{2}$. As a comparison, other
theoretical works and available experimental data are listed.}%
\begin{ruledtabular}
\begin{tabular}{ccccccc}
Property&This study&Previous calculation&Experiment\\
\hline
&VASP-GGA&FP-LTMO-GGA& \\
\emph{a$_{0}$} ({\AA})& 5.62 & 5.61$^{\emph{a}}$ & 5.60001(3)$^{\emph{b}}$, 5.598(3)$^{\emph{d}}$ \\
\emph{C$_{11}$} (GPa)& 349.5 & 376.0$^{\emph{a}}$ & 367$^{\emph{c}}$ \\
\emph{C$_{12}$} (GPa)& 111.4 & 109.8$^{\emph{a}}$ & 106$^{\emph{c}}$ \\
\emph{C$_{44}$} (GPa)& 70.6 & 68.1$^{\emph{a}}$ & 79$^{\emph{c}}$ \\
\emph{B} (GPa)&191 &198$^{\emph{a}}$, 198$^{\emph{d}}$ &198(2)$^{\emph{b}}$, 195(2)$^{\emph{d}}$ \\
\emph{B$^{'}$} &4.5 &4.5$^{\emph{a}}$, 4.2$^{\emph{d}}$ &4.6(3)$^{\emph{b}}$, 5.4(2)$^{\emph{d}}$ \\
\emph{G} (GPa)&87.1  &94.1$^{\emph{a}}$  &97$^{\emph{c}}$  \\
\emph{E} (GPa)&226.8  &243.8$^{\emph{a}}$  &249$^{\emph{c}}$ \\
$\upsilon$ & 0.302 &0.295$^{\emph{a}}$  & 0.285$^{\emph{c}}$ \\
Hardness (GPa)&27.33 & & \\
\emph{E$_{g}$} (eV)&4.673&4.522$^{\emph{d}}$&6$^{\emph{e}}$\\
\end{tabular}
\end{ruledtabular}
\label{mechanical} $^{\emph{a}}$ Reference \cite{Kanchana},
$^{\emph{b}}$ Reference \cite{Idiri}, $^{\emph{c}}$ Reference
\cite{Macedo}, $^{\emph{d}}$ Reference \cite{Staun},
$^{\emph{e}}$Reference \cite{Sviridova}.
\end{table*}

At ambient condition thorium dioxide crystallize in a CaF$_{2}$-like
ionic structure. Its cubic unit cell is composed of four ThO$_{2}$
formula units with the thorium atoms and the oxygen atoms in
4\emph{a} and 8\emph{c} sites, respectively [see Fig. 1(a)]. Each Th
atom is surrounded by eight O atoms forming a cube and each O
connects with four Th atoms to build a tetragon. A particular
feature of this structure is the presence of a large octahedral hole
sited at position $\left( \frac{1}{2},\frac{1}{2},\frac{1}{2}\right)
$. The present optimized lattice constant ($a_{0}$) is 5.62 \AA (see
Table I), in good agreement with the experimental \cite{Idiri,Staun}
value of 5.6 \AA.

Elastic constants can measure the resistance and mechanical features
of crystal to external stress or pressure, thus describing the
stability of crystals against elastic deformation. For small strains
Hooke's law is valid and the crystal energy $E$ is a quadratic
function of strain \cite{Nye1}.Thus, to obtain the total minimum
energy for calculating the elastic constants to second order, a
crystal is strained and all the internal parameters relaxed.
Consider a symmetric $3\times3$ nonrotating strain tensor
$\mathbf{\varepsilon}$ which has matrix elements $\varepsilon_{ij}$
($i,j=1,2,$ and $3$) defined by
\begin{eqnarray}
\varepsilon_{ij}=\left(
\begin{array}
[c]{ccc}%
e_{1} & \frac{e_{6}}{2} & \frac{e_{5}}{2}\\
\frac{e_{6}}{2} & e_{2} & \frac{e_{4}}{2}\\
\frac{e_{5}}{2} & \frac{e_{4}}{2} & e_{3}%
\end{array}
\right)  .
\end{eqnarray}
Such a strain transforms the three lattice
vectors defining the unstrained Bravais lattice $\left\{
\mathbf{a}_{k},k=1,2,\text{and }3\right\}  $ to the strained vectors
\cite{Alou} $\left\{ \mathbf{a}_{k}^{\prime}\right\}  $ as
defined by%
\begin{eqnarray}
\mathbf{a}_{k}^{\prime}=(\mathbf{I}+\mathbf{\varepsilon})\mathbf{a}_{k},
\end{eqnarray}
where $\mathbf{I}$ is a unit $3\times3$ matrix. Each lattice vector
$\mathbf{a}_{k}$ or $\mathbf{a}_{k}^{\prime}$ is a $3\times1$
matrix. The change in total energy due to above strain (1) is
\cite{Nye1}
\begin{align}
\Delta E&=E(V,\{\epsilon_{i}\})-E(V,0)\nonumber\\
&=V\sum_{i=1}^{6}\sigma_{i}e_{i}+\frac{V}{2}\sum_{i,j=1}^{6}C_{ij}e_{i}e_{j}+O(\{e_{i}^{3}\}),
\end{align}
where $E(V,0)$ is the total energy for the unstrained crystal,
$\sigma_{i}$ are the elements of stress tensor, and $C_{ij}$ are the
elastic constants. For cubic structure of ThO$_{2}$, there are three
independent elastic constants, i.e., $C_{11}$, $C_{12}$, and
$C_{44}$, which are calculated through a proper choice of the set of
strains $\{e_{i},i=1,...,6\}$ listed in Table II. In our
first-principles calculations, the strain amplitude $\delta$ is
varied in steps of 0.006 from $\delta$=-0.036 to 0.036 and the total
energies $E(V,\delta)$ at these strain steps are calculated, and
then fitted through the strains with the corresponding parabolic
equations of $\Delta E/V$ as given in Tables II to yield the
required second-order elastic constants. Note that while computing
these energies all atoms are allowed to relax with the cell shape
and volume fixed by the choice of strains $\{e_{i}\}$. Our
calculated elastic constants for the ground-state fluorite structure
of ThO$_{2}$ are listed in Table \ref{mechanical}. As a comparison,
previous theoretical results based on the full-potential linear
muffin-tin orbital (FP-LMTO) method \cite{Kanchana}, as well as
available experimental data\cite{Macedo} are also presented in Table
I. It is clear that our calculated elastic constants are in good
agreement with the experimental results.

\begin{table}
\caption{ Three strain combinations in the strain tensor [Eq. (1)]
to
calculate the three elastic constants of cubic ThO$_{2}$.}%
\begin{ruledtabular}
\begin{tabular}{ccccccc}
Strain&Parameters (unlisted \emph{e$_{i}$}=0)&\emph{$\Delta$E/V} in \emph{O($\delta$$^{2}$)}\\
\hline \emph{$\epsilon$$^{1}$}& \emph{e$_{1}$}=$\delta$
\emph{e$_{2}$}=$\delta$ \emph{e$_{3}$}=$\delta$&
$\frac{3}{2}$$(C_{11}+2C_{12})\delta^{2}$\\
\emph{$\epsilon$$^{2}$}& \emph{e$_{1}$}=$\delta$
\emph{e$_{3}$}=$\delta$&
$(C_{11}+C_{12})\delta^{2}$\\
\emph{$\epsilon$$^{3}$}& \emph{e$_{4}$}=$\delta$
\emph{e$_{5}$}=$\delta$ \emph{e$_{6}$}=$\delta$&
$\frac{3}{2}$$C_{44}\delta^{2}$\\
\end{tabular}
\label{strain}
\end{ruledtabular}
\end{table}

After obtaining elastic constants, we can calculate bulk and shear
moduli from the Voigt-Reuss-Hill (VRH) approximations
\cite{Voigt,Reuss,Hill}. The Voigt bounds \cite{Voigt,Watt} on the
effective bulk modulus \emph{B$_{V}$} and shear modulus
\emph{G$_{V}$} are
\begin{eqnarray}
B_{V}=(C_{11}+2C_{12})/3
\end{eqnarray}
and
\begin{eqnarray}
G_{V}=(C_{11}-C_{12}+3C_{44})/5.
\end{eqnarray}
Under Reuss approximation \cite{Reuss}, the Reuss bulk modulus
\emph{B$_{R}$} and Reuss shear mudulus \emph{G$_{R}$} are
\begin{eqnarray}
B_{R}=B_{V}
\end{eqnarray}
and
\begin{eqnarray}
G_{R}=5(C_{11}-C_{12})C_{44}/[4C_{44}+3(C_{11}-C_{12})].
\end{eqnarray}
The bulk modulus \emph{B} and shear modulus \emph{G}, based on Hill
approximation \cite{Hill}, are arithmetic average of Voigt and Reuss
elastic modulus, i.e., \emph{B}=$\frac{1}{2}(B_{R}+B_{V})$ and
\emph{G}=$\frac{1}{2}(G_{R}+G_{V})$. The Young's modulus \emph{E}
and Poisson's ratio $\upsilon$ for an isotropic material are given
by\cite{Ravindran}
\begin{eqnarray}
E=\frac{9BG}{3B+G},
\end{eqnarray}
and
\begin{eqnarray}
\upsilon=\frac{3B-2G}{2(3B+G)}.
\end{eqnarray}

The calculated results for these moduli and Poisson's ratio for the
fluorite ThO$_{2}$ are listed in Table I. Note that we have also
calculated the bulk modulus \emph{B} by fitting the Murnaghan
equation of state. The derived bulk modulus turns out to be exactly
the same as that from the above VRH approximation, which again
indicates that our calculations are consistent and reliable. For
comparison, other theoretical results \cite{Kanchana, Staun} and
available experimental data\cite{Idiri, Macedo, Staun} are also
shown in Table I. It can be seen that on the whole, our present
results compare well with those previous experimental and FP-LMTO
theoretical results. Concerning the Poisson's ratio, it is well
known that for the common materials that have much smaller shear
moduli compared with the bulk moduli, their Poisson's ratio is close
to 1/3. For the present ThO$_{2}$ system, our calculated shear
modulus $G$ is much lower than the bulk modulus $G$. Thus, our
calculated result of 0.302 for the Poisson's ratio, as well as the
derived result of 0.285 according to the experimentally determined
elastic constants \cite{Macedo} and using Eq. (9), can be well
understandable.

Furthermore, hardness is also one fundamental physical quantity when
considering the phase stability and mechanical properties. Note that
the hardness is important for the applications of ThO$_{2}$ as both
nuclear material and optical components. So here we calculate the
hardness of ThO$_{2}$ by using the approach raised by Simunek
\emph{et al.} \cite{SimunekPRL}. In the case of two atoms 1 and 2
forming one bond of strength \emph{s$_{12}$} in a unit cell of
volume $\Omega$, the expression for hardness has the form
\cite{SimunekPRL}
\begin{equation}
H=(C/\Omega)b_{12}s_{12}e^{-\sigma\!f_{2}},
\end{equation}
where
\begin{equation}
s_{12}=\sqrt{(e_{1}e_{2})}/(n_{1}n_{2}d_{12}), e_{i}=Z_{i}/r_{i}
\end{equation}
and
\begin{equation}
f_{2}=(\frac{e_{1}-e_{2}}{e_{1}+e_{2}})^{2}=1-[\sqrt{(e_{1}e_{2})}/(e_{1}+e_{2})]^{2}
\end{equation}
are the strength and ionicity of the chemical bond, respectively,
and \emph{d$_{12}$} is the interatomic distance; \emph{C}=1550 and
$\sigma$=4 are constants. The radius \emph{r$_{i}$} is chosen to
make sure that the sphere centered at atoms \emph{i} in a crystal
comtains exactly the valence electronic charge \emph{Z$_{i}$}. For
fluorite structure ThO$_{2}$, \emph{b$_{12}$}=32 counts the
interatomic bonds between atoms Th (1) and O (2) in the unit cell,
\emph{n$_{1}$}=8 and \emph{n$_{2}$}=4 are coordination numbers of
atom Th and O, respectively, \emph{r$_{1}$}=1.81 ({\AA}) and
\emph{r$_{2}$}=1.00 ({\AA}) are the atomic radii for Th and O atoms,
respectively, \emph{Z$_{1}$}=12 and \emph{Z$_{2}$}=6 are valence
charge for Th and O atoms, respectively, \emph{d$_{12}$}=2.43
({\AA}) is the interatomic distance, and $\Omega$=145.53
({\AA}$^{3}$) is the volume of unit cell. Using Eqs. (10)-(12), we
obtain \emph{s$_{12}$}=0.081 and \emph{f$_{2}$}=0.0025. The hardness
of ThO$_{2}$ at its ground-state fluorite structure is thus given by
\emph{H}=27.3 (GPa). This indicates that the fluorite ThO$_{2}$ is a
hard material and approaches to a superhard material (hardness $>$
40 GPa). The high hardness of this crystal can be understood from
the dense crystal structure, which results in high valence electron
density and short bond distances. The unusual mixture of covalent
and ionic components in the Th-O chemical bond will be discussed in the following
subsection.

\subsection{Electronic structure and charge distribution}%

\begin{figure}[tbp]
\begin{center}
\includegraphics[width=1.0\linewidth]{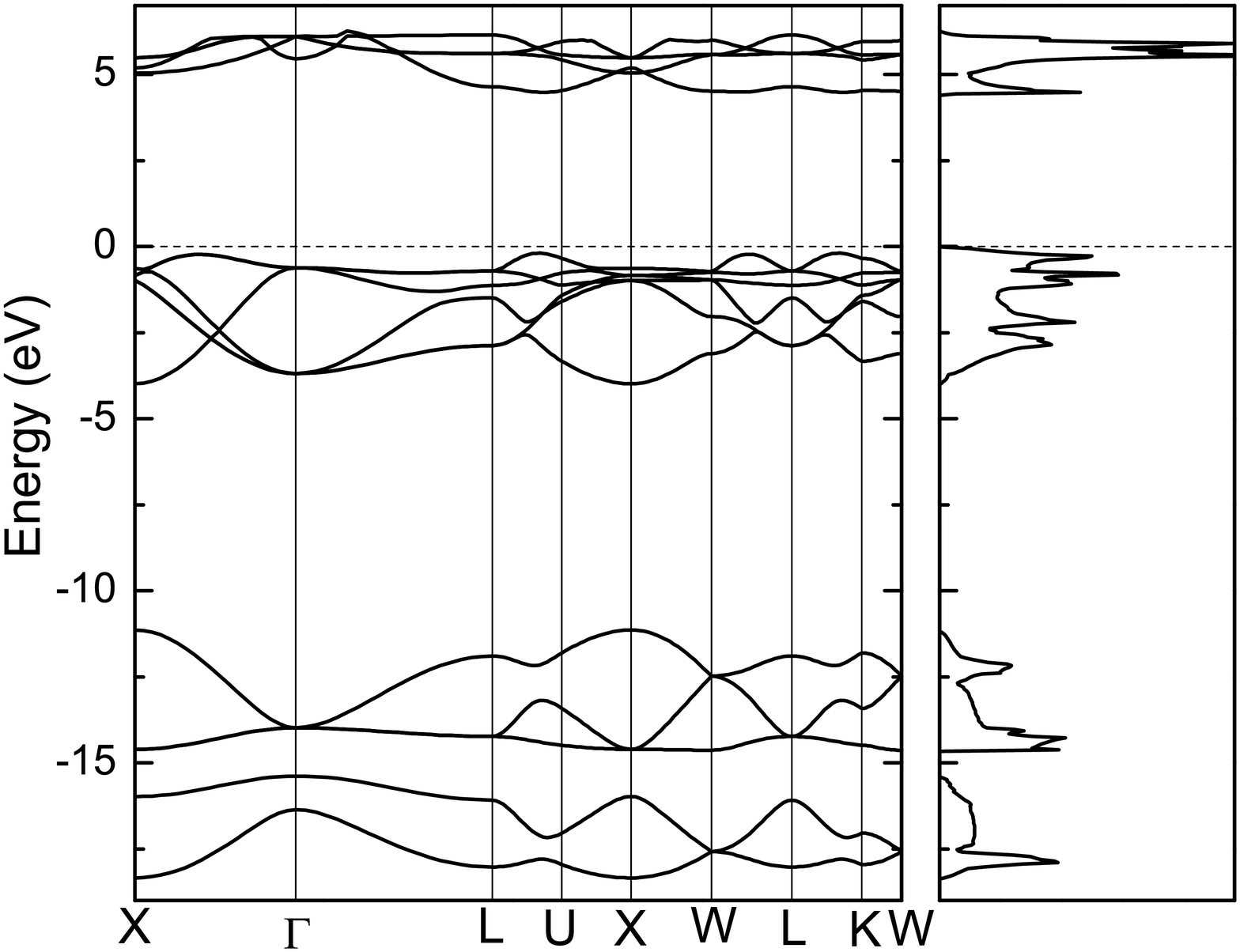}
\end{center}
\caption{Band structure (left panel) and total density of states
(DOS, right panel) for ground state ThO$_{2}$ with Fermi energy
level \emph{E$_{F}$} taken at 0 eV as shown by the dashed lines.}
\label{band}
\end{figure}

\begin{figure}[tbp]
\begin{center}
\includegraphics[width=1.0\linewidth]{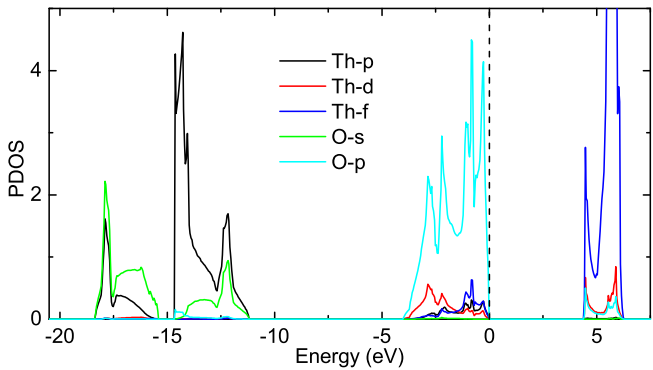}
\end{center}
\caption{(Color online) Partial density of states (PDOS) for ground
state ThO$_{2}$ at 0 GPa. The Fermi energy level is zero.}
\label{PDOS}
\end{figure}

Almost all the macroscopical properties of materials, such as
hardness, elasticity, and conductivity, originate from their
electronic structure properties as well as chemical bonding nature.
Therefore, it is necessary to perform the electronic structure
analysis of ThO$_{2}$. The calculated band structure (left panel)
and total density of states (DOS, right panel) of fluorite ThO$_{2}$
are shown in Fig. \ref{band}. The present calculated energy band gap
of 4.673 eV (listed in Table \ref{mechanical}), which is consistent
with previous FT-LMTO theoretical result of 4.522 eV \cite{Staun},
indicates that ThO$_{2}$ is a typical insulator. The underestimation
of band gap compared with experimental value of 6 eV
\cite{Sviridova} is due to the drawback of the exchange-correlation
approximation (GGA). As a comparison, the orbital-resolved partial
densities of states (PDOS) for one Th atom and two O atoms in a unit
cell are also presented in Fig. \ref{PDOS}. One can see that the low
bands covering from -18.33 to -11.22 eV mainly consist of O $2s$ and
Th $6p$ states which shows a clear hybridization. This
hybridization is also responsible for the covalency of ThO$_{2}$ (see further discussion below).
The high valence bands located just below the Fermi level are mainly
contributed by O $2p$ states with a little Th 6\emph{d} and
5\emph{f} states and have a band width of 3.79 eV. This illustrates
a significant charge transfer from Th 6\emph{d} and 5\emph{f} states
to O 2\emph{p} states. As for the conduction bands, the DOS is
mainly featured by Th $5f$ states, mixed with a little Th 6\emph{d}
and O 2\emph{p} states and has a width of 1.78 eV.

\begin{figure}[tbp]
\begin{center}
\includegraphics[width=1.0\linewidth]{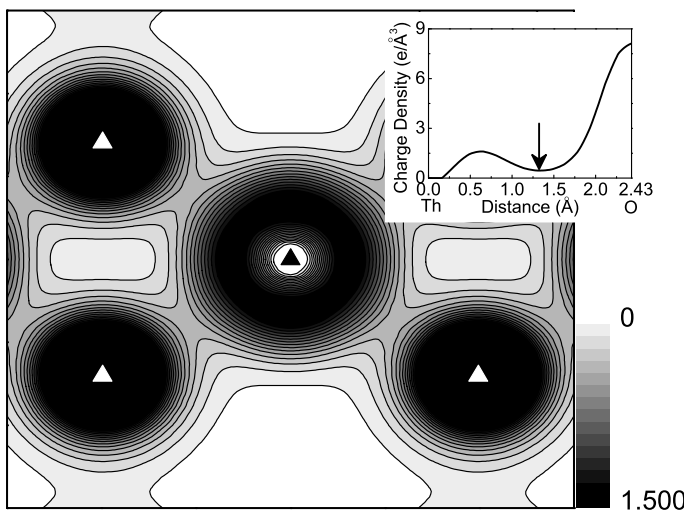}
\end{center}
\caption{Valence charge density of ThO$_{2}$ in (1$\bar{1}$0) plane.
There $\blacktriangle$ stands for Th, $\vartriangle$ for O atoms.
The contour lines are drawn from 0.0 to 1.5 at 0.1 e/{\AA}$^{3}$
intervals. The line charge density distribution between Th atom and
the nearest neighbor O atom is shown in the inset, where the arrow
indicates the minimum value.} \label{charge}
\end{figure}%

In order to gain more insight into the bonding nature of ground
state ThO$_{2}$, we also investigate the valence charge density
distribution. The calculated valence charge density map of the
(1$\bar{1}$0) plane is plotted in Fig. \ref{charge}. It is obvious
from Fig. 4 that the charge density around Th and O ions are all
near spherical distribution with slightly deformed toward the
direction to their nearest neighboring atoms. Furthermore, the
charge density around Th and O ions is high while there is almost no
valence charge in the large octahedral-hole interstitial region.
This suggests that remarkable ionization exists for thorium and
oxygen ions and significant insulating property exhibits in
ThO$_{2}$. In order to describe the ionicity quantitatively and more
clearly, we plot in the inset in Fig. \ref{charge}  the line charge
density distribution along the nearest Th-O bond. Therein the arrow
indicates the charge density minimum of 0.45 e/{\AA}$^{3}$. The
distance between this charge density minimum point and Th ion is
1.32 \AA. According to this, we can define the ionic radius of Th in
ThO$_{2}$ as $r_{\text{Th}}$=1.32 \AA. After subtracting this value
from the Th-O bond length (2.43 \AA), one obtains the oxygen ionic
radius in ThO$_{2}$ as $r_{\text{O}}$=1.11 \AA. The corresponding
valency of Th and O in ThO$_{2}$ can be estimated by calculating the
valence charges within the spheres of ionic radii. After
integration, we find that there are 8.166 electrons around Th and
6.452 electrons around O. As a result, the valency of Th and O can
be formally represented as Th$^{3.834+}$ and O$^{0.452-}$,
respectively. This indicates that the Th and O atoms in ThO$_{2}$
are ionized inequivalently. In other words, Th is ionized almost as
Th$^{4+}$, while O is weakly ionized. This fact, on the other hand,
suggests that besides ionicity, the Th-O bond also includes a
prominent covalent components, which is consistent with the above
PDOS analysis. In fact, the  minimum value 0.45 e/{\AA}$^{3}$ of
charge density along the Th-O bond in ThO$_{2}$ is prominently
higher than that in typical ionic crystals. Based on the PDOS and
charge density analysis, therefore, we derive that the Th-O bond is
a mixture of covalent and ionic components. From this aspect, it is
now understandable that the previous full-ionic model \cite{Harding}
has largely underestimated the bulk modulus of ThO$_{2}$.

\subsection{Phonon dispersion curve}%
\begin{figure}[tbp]
\begin{center}
\includegraphics[width=1.0\linewidth]{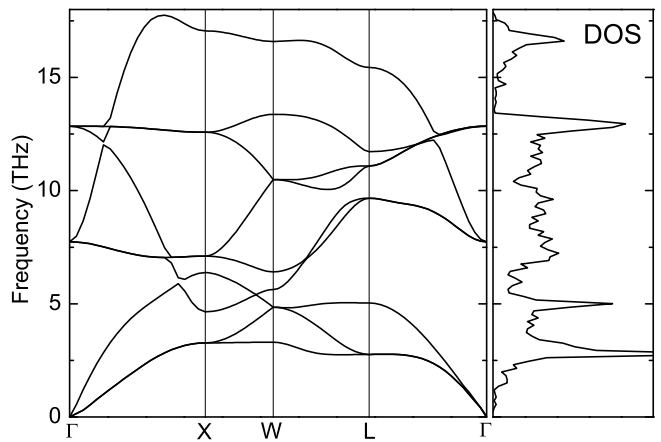}
\end{center}
\caption{Calculated phonon dispersion curves along the high-symmetry
directions (left panel) and total density of states (DOS, right
panel) for ground state ThO$_{2}$.} \label{phonon}
\end{figure}%

The calculation of phonon frequencies of the crystalline structure
is one of the basic aspects when considering the phase stability,
phase transformations, and thermodynamics of crystalline materials.
By using the Hellmann-Feynman theorem and the direct method, we have
calculated the phonon curves along some high-symmetry directions in
the Brillouin zone, together with the phonon density of states. For
the phonon dispersion calculation, we use the 2$\times$2$\times$2
fcc supercell containing 96 atoms and the 4$\times$4$\times$4
Monkhorst-Pack \emph{k}-point mesh for the Brillouin zone
integration. In order to calculate the Hellmann-Feynman forces, we
displace four atoms (two Th and two O atoms) from their equilibrium
positions and the amplitude of all the displacements is 0.03 \AA.
The calculated phonon dispersion curves along the
$\Gamma$-$X$-$W$-$L$-$\Gamma$ directions are displayed in Fig.
\ref{phonon}. The $\Gamma$-$X$, $X$-$W$, and $L$-$\Gamma$ lines are
along $<$0 0 1$>$, $<$1 0 2$>$, and $<$1 1 1$>$ directions,
respectively.

For ThO$_{2}$ in the CaF$_{2}$-structure primitive cell, there are
only three atoms. Therefore, nine phonon modes exist in the
dispersion relations. One can see that there is no gap between the
optic modes and the acoustic branches and the LO-TO also has no
splitting at $\Gamma$ point. Due to the fact that thorium is heavier
than oxygen atom, the vibration frequency of thorium atom is lower
than that of oxygen atom. Therefore, the phonon density of states
can be viewed as two parts. One is the part lower than 5.5 THz where
the main contribution comes from the thorium sublattice, while the
other part higher than 5.5 THz, included both acoustic and optical
branches, are dominated by the dynamics of the light oxygen atoms.

\section{High pressure behavior of ThO$_{2}$}

\begin{table*}[ptb]
\caption{Calculated elastic constants, bulk modulus
\emph{B}, pressure derivative of the bulk
modulus \emph{B$^{'}$}, shear modulus \emph{G}, Young's modulus
\emph{E}, Poisson's ratio $\upsilon$ for cotunnite-type ThO$_{2}$ at
0 GPa. Except the Poisson's ratio, all other values are in units of
GPa.}
\begin{ruledtabular}
\begin{tabular}{cccccccccccccccc}
\emph{C$_{11}$}&\emph{C$_{22}$}&\emph{C$_{33}$}&\emph{C$_{44}$}&\emph{C$_{55}$}&\emph{C$_{66}$}&\emph{C$_{12}$}&\emph{C$_{23}$}&\emph{C$_{13}$}&\emph{B}&\emph{B$^{'}$}&\emph{G}&\emph{E}&$\upsilon$\\
\hline
299.3 & 256.3  & 235.6 & 37.4  & 54.9 &84.9&73.7&95.5&104.8&148&7.8&65.9&172.2&0.307\\
\end{tabular}
\label{Pnmaelastic}
\end{ruledtabular}
\end{table*}

In order to investigate the high pressure behavior of ThO$_{2}$, we
have optimized the structural parameters of its \emph{Fm$\bar{3}$m}
and \emph{Pnma} phases at different pressures by using GGA method.
To aviod the Pulay stress problem, we perform the structure
relaxation calculations at fixed volumes rather than constant
pressures. For \emph{Fm$\bar{3}$m} phase, due to its high symmetry,
the structure relaxation calculations are performed at fixed volumes
with no relaxation of coordinates. However, for \emph{Pnma} phase,
the coordinates of atoms and the cell shape are necessary to be
optimized due to their internal degrees of freedom.

The optimized structural lattice parameters \emph{a}, \emph{b}, and
\emph{c} for the \emph{Pnma} phase at 0 GPa are 6.174, 3.776, and
7.161 {\AA}, respectively, giving \emph{V}=167.0 {\AA}$^{3}$. This
volume is prominently smaller than the equilibrium volume of 177.5
{\AA}$^{3}$ for the \emph{Fm$\bar{3}$m} phase. Using the same method
as that has been employed in our previous work on orthorhombic
BeH$_{2}$ \cite{Wang}, we have applied the strains and calculated
the elastic constants, various moduli, and Poisson's ratio
$\upsilon$ for cotunnite-type ThO$_{2}$ at 0 GPa. The results are
collected in Table \ref{Pnmaelastic}. From Table \ref{Pnmaelastic}
the following prominent features can be seen: (i) The orthorhombic
ThO$_{2}$ at zero pressure is mechanically stable because its
elastic constants satisfy the following mechanical stability
criteria \cite{Nye2}.
\begin{align}
&  C_{11}>0,C_{22}>0,C_{33}>0,C_{44}>0,C_{55}>0,C_{66}>0,\nonumber\\
&  [C_{11}+C_{22}+C_{33}+2(C_{12}+C_{13}+C_{23})]>0,\nonumber\\
&  (C_{11}+C_{22}-2C_{12})>0,(C_{11}+C_{33}-2C_{13})>0,\nonumber\\
&  (C_{22}+C_{33}-2C_{23})>0.
\end{align}
In fact, one can see from Table \ref{Pnmaelastic} that the
calculated $C_{12}$, $C_{23}$, and $C_{13}$ are largely smaller than
$C_{11}$, $C_{22}$, and $C_{33}$. Therefore, the mechanical
stability criteria is easily satisfied; (ii) Although the
equilibrium volume of cotunnite-type ThO$_{2}$ is distinctly
compressed, its bulk modulus, shear modulus, and Young's modulus are
all smaller by approximately 25\% than those in fluorite-type
ThO$_{2}$. This intriguing behavior has also been observed in the
UO$_{2}$ system \cite{Geng}, which, like ThO$_{2}$, will undergo a
$Fm\bar{3}m\mathtt{\rightarrow}Pnma$ phase transition. Since the
$5f$ orbital band is occupied in UO$_{2}$, while it is almost empty in
ThO$_{2}$, thus this similarity of the softening in moduli upon
phase transition for the two systems is clearly unrelated with the
$5f$ orbitals. From this aspect, we speculate that most actinide
dioxides share this phase-transition involved similarity. Again, we have confirmed for the \emph{Pnma} phase
that the bulk modulus \emph{B} calculated by fitting the Murnaghan
equation of state equals to that through VRH approximation; (iii)
The Poisson's ratio $\upsilon$ of ThO$_{2}$ in $Pnma$ phase is
nearly the same as in $Fm\bar{3}m$ phase, i.e., close to 1/3. This
is understandable since the shear modulus is much smaller than the
bulk modulus in the both two phases.

\begin{figure}[tbp]
\begin{center}
\includegraphics[width=1.0\linewidth]{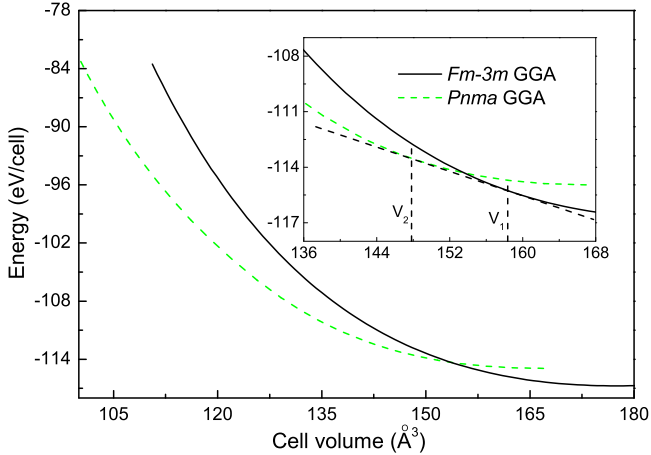}
\end{center}
\caption{(Color online) Comparison of total energy vs the cell
volume for ThO$_{2}$ in \emph{Fm$\bar{3}$m} and \emph{Pnma} phases.
A phase transition at 26.5 GPa is predicted by the slope of the
common tangent rule, as shown in the inset.} \label{energy}
\end{figure}

\begin{figure}[tbp]
\begin{center}
\includegraphics[width=1.0\linewidth]{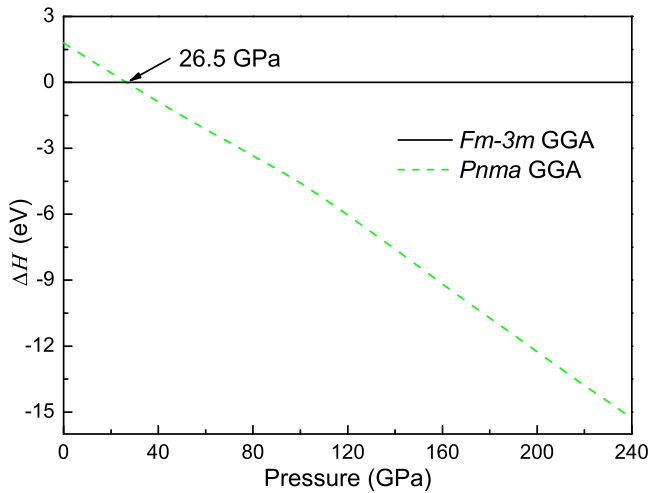}
\end{center}
\caption{(Color online) Calculated enthalpy differences of
\emph{Pnma} phase with respect to \emph{Fm$\bar{3}$m} phase as a
function of pressure.} \label{enthalpy}
\end{figure}

Now let us see the phase transition energetics of ThO$_{2}$. The
total energies (per unit cell) of the two phases at different
volumes are calculated and shown in Fig. \ref{energy}. Obviously,
the \emph{Fm$\bar{3}$m} phase is stable under ambient pressure while
under high pressure the \emph{Pnma} phase becomes stable. According
to the rule of common tangent of two energy curves, a phase
transition at 26.5 GPa is predicted by the slope shown in the inset
of Fig. \ref{energy}. Besides, we also determine the phase
transition pressure by comparing the Gibbs free energy as a function
of pressure. At 0 K, the Gibbs free energy is equal to the enthalpy
\emph{H}, expressed as \emph{H}=\emph{E}+\emph{PV}. Figure
\ref{enthalpy} shows the relative enthalpies of the cotunnite phase
phase with respect to the fluorite phase as a function of pressure.
The crossing between the two enthalpy curves in Fig. \ref{enthalpy}
readily gives phase transition pressure of 26.5 GPa, which is fully
consistent with the above result in terms of the common tangent
rule. This value is well close to the experiment measurement by
Idiri \emph{et al}. \cite{Idiri} and and by Jayaraman \emph{et al}.
\cite{Jaya} as $\mathtt{\sim}$30 GPa. Here, the minor theoretical
underestimation by $\mathtt{\sim}$5.0 GPa is speculated to be caused
by the possible existence of an energy barrier with an amplitude of
$\Delta w$. To overcome this energy barrier, the external pressure
$P^{\prime}$ should be larger than the conventional
common-tangent-rule determined pressure $P$ by an amount $\Delta
P$=$P^{\prime}\mathtt{-}P$. According to the experimentally measured
transition pressure $P^{\prime}$ \cite{Idiri} and our theoretically
obtained common tangent curve in Fig. \ref{energy}, we deduce the
energy barrier amplitude to be $\mathtt{\sim}$0.06 eV (per formula
unit). This value is too small to survive the \emph{Pnma} phase to
ambient conditions, which is consistent with the fact that to date,
the \emph{Pnma} phase of ThO$_{2}$ has only be observed under high
pressures.

\begin{figure}[tbp]
\begin{center}
\includegraphics[width=1.0\linewidth]{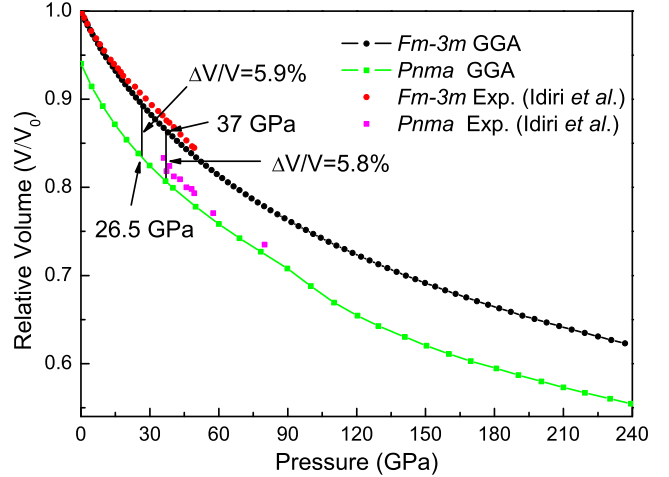}
\end{center}
\caption{(Color online) Calculated compression curves of ThO$_{2}$
compared with experimental measurements. The volume collapses at our
predicted phase transition point 26.5 GPa and experimental phase
transition pressure 37 GPa are labeled.} \label{pressure}
\end{figure}

\begin{figure}[tbp]
\begin{center}
\includegraphics[width=1.0\linewidth]{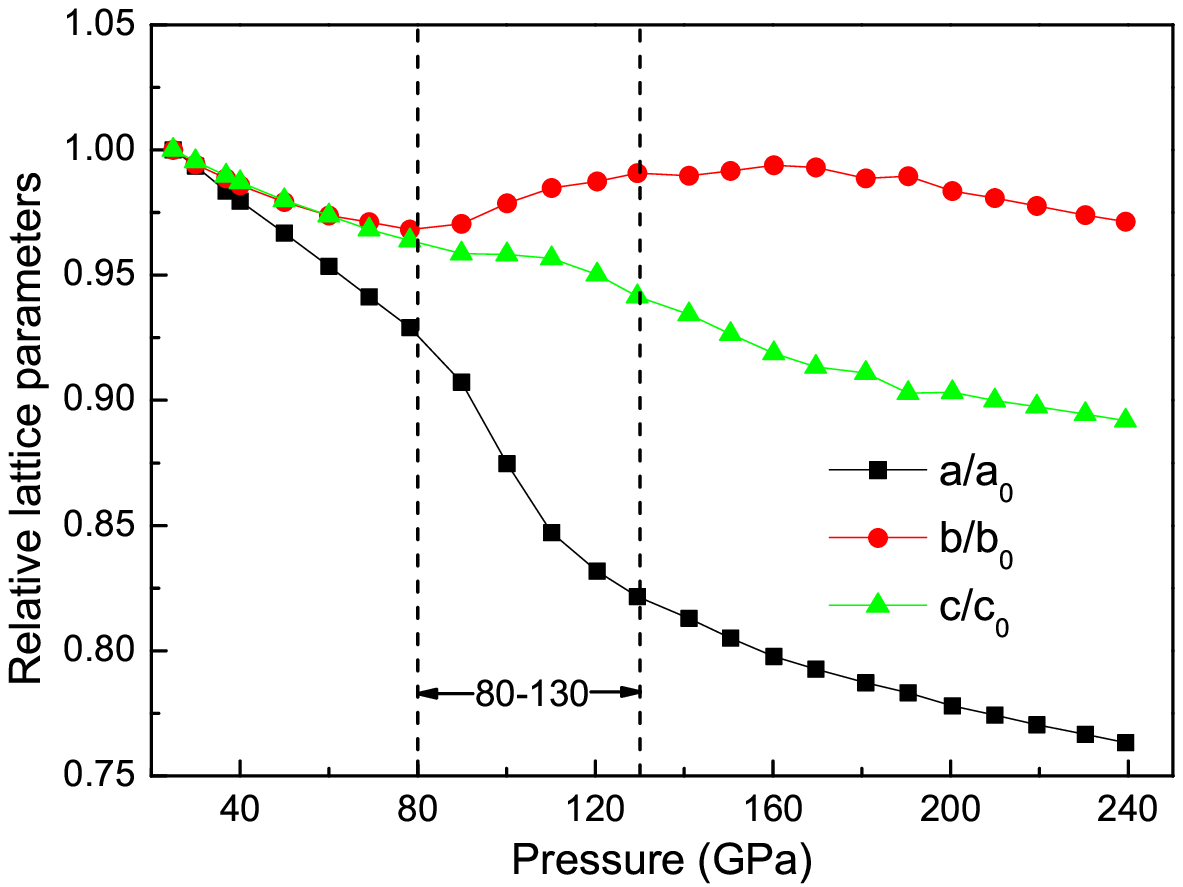}
\end{center}
\caption{(Color online) Pressure behavior of the relative lattice
parameters of the \emph{Pnma} phase, where the drastic change in the
relative lattice constants (region between dashed lines) indicates
an isostructural transition.} \label{lattice}
\end{figure}

Figure \ref{pressure} shows the relative volume \emph{V/V$_{0}$}
evolution with pressure for ThO$_{2}$ in both  \emph{Fm$\bar{3}$m}
and \emph{Pnma} phases. For comparison, the experimental data are
also shown in the figure. Clearly, our calculated P-V equation of
state is well consistent with the experimental measurement for the
both two phases of ThO$_{2}$. Specially, at the calculated
transition pressure (26.5 GPa), our result in Fig. \ref{pressure}
gives that the volume collapse upon phase transition is 5.9\%. This
value is very close to the experimental data of 6.1\% \cite{Idiri}
with a little bit underestimation, which is obviously within the
accuracy of GGA. Note that unlike the cubic \emph{Fm$\bar{3}$m}
phase, in which all coordinated are completely determined at each
volume due to the symmetry, the orthorhombic \emph{Pnma} phase at
each volume has two additional internal degrees of freedom that must
be fully relaxed to obtain the energy minimum. Thus, we have further
calculated and plotted in Fig. \ref{lattice} the pressure dependence
of the three lattice parameters (with respect to their equilibrium
values) for the \emph{Pnma} phase of ThO$_{2}$. In the
experimentally reported pressure range from $\mathtt{\sim}$30 GPa to
$\mathtt{\sim}$80 GPa, our calculated evolution of the relative
lattice parameters in the \emph{Pnma} phase are well consistent with
the experimental observation \cite{Idiri}. In this pressure region,
one can see from Fig. \ref{lattice} that the responses of the three
relative lattice parameters to the compression are anisotropic in
the sense that the compression of the middle axis $a$ is most rapid
compared to those of the long axis $c$ and small axis $b$, which
vary upon compression almost in the same tendency. When the pressure
becomes higher to be between 80 and 130 GPa, remarkably, it reveals
in Fig. \ref{lattice} that all the three relative lattice parameters
undergo dramatic variations by the fact that the small axis $b$ has
a strong rebound and the middle $a$ is collapsed. When the pressure
is beyond 130 GPa, then the variations of the three relative lattice
parameters become smooth and approach isotropic compression. This
signifies a typical isostructural transition for the \emph{Pnma}
phase of ThO$_{2}$. It should be stressed that this isostructural
transition is not unique for ThO$_{2}$. Similar phenomenon has also
been observed in other actinide dioxides \cite{Geng}.

\begin{figure}[tbp]
\begin{center}
\includegraphics[width=1.0\linewidth]{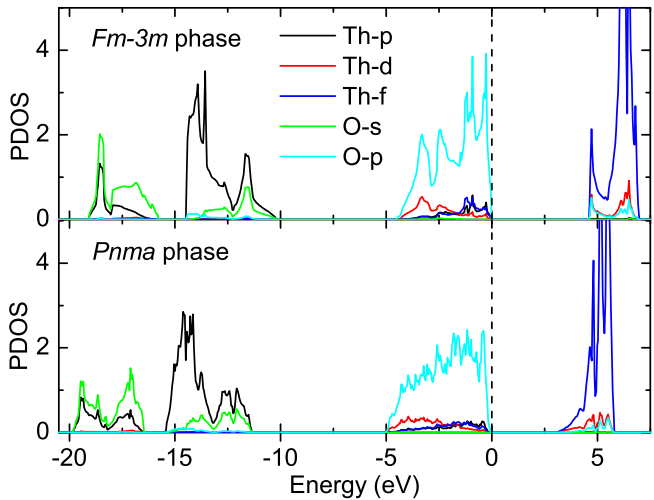}
\end{center}
\caption{(Color online) Partial density of states (PDOS) for
\emph{Fm$\bar{3}$m} phase (a) and \emph{Pnma} phase (b) both at 26.5
GPa. The Fermi energy level is zero. The energy gaps of
\emph{Fm$\bar{3}$m} phase and \emph{Pnma} phase are 4.584 eV and
3.153 eV, respectively.} \label{twoDOS}
\end{figure}

To see the change of electronic structure of ThO$_{2}$ under high
pressure, we present in Fig. \ref{twoDOS} the orbital-resolved PDOS
for the \emph{Fm$\bar{3}$m} and \emph{Pnma} phases at the
theoretical transition pressure of 26.5 GPa. One can see that there
occurs an understandable narrowing in the band gap (4.6 eV for the
\emph{Fm$\bar{3}$m} phase and 3.15 eV for the \emph{Pnma} phase).
The valence and conduction bands carry a sizeable downward shift to
stabilize the high-pressure \emph{Pnma} phase. In addition, the
valence and conduction bands are more widened and smooth in the
\emph{Pnma} phase. However, these changes in the electronic
structure are trivial and thus cannot be associated with the exotic
transition-accompanied volume collapse in ThO$_{2}$ and other
actinide dioxides. Concerning the 5$f$ electron effect, here we
would like to present our viewpoint as follows. First, the 5$f$
orbital is almost empty for both two phases of ThO$_{2}$. Therefore,
the transition-accompanied volume collapse in ThO$_{2}$ is
irrelevant to the 5$f$ orbital effect. Second, In other actinide
dioxides, such as PuO$_{2}$, although taking into account the 5$f$
electronic interaction is crucial to obtain more reasonable ground
state \cite{Zhang}, this local correlation effect is not expected to
play a role in explaining the transition-accompanied volume collapse
in these 5$f$-occupied actinide dioxides. The reason is simply
because that inclusion of 5$f$ electronic localization will weaken
the binding of cations and anions, causing an increase instead of a
decrease in the volume of system. Therefore, the phenomenon of
volume collapse during high-pressure phase transition of the
actinide dioxides is mainly attributed to the ionic (instead of
electronic) response to the external compression.

\section{CONCLUSION}

In summary, the ground state properties as well as the high pressure behavior of ThO$_{2}$ were studied
by means of the first-principles DFT-GGA method. The elastic constants and their derived moduli and Poisson's ratio were calculated
for both the ambient \emph{Fm$\bar{3}$m} and the high-pressure \emph{Pnma} phases of ThO$_{2}$ and were shown to accord well with
recent reliable experiments. Based on these results, mechanical stability of the \emph{Pnma} phase at zero pressure was predicted. The
hardness, lattice vibration dynamics, and the Th-O chemical bond of the ambient phase were calculated and analyzed in order to support
the practical application of of ThO$_{2}$. We showed that the Th-O bond displays a mixed ionic/covalent character, with the valence
of Th and O ions represented as Th$^{3.834+}$ and O$^{0.452-}$. Here the ionicity is
mainly featured by charge transfer from Th $6d/5f$ states to O $2p$
states, while the covalency is manifested by hybridization
of oxygen 2\emph{s} and thorium 6\emph{p} states.
This mixed ionic/covalent feature makes ThO$_{2}$ a hard material, with its hardness calculated
in this paper to be $\mathtt{\sim}$27 GPa. As another main task, we studied phase transition of ThO$_{2}$ under high pressure.
Our calculated \emph{Fm$\bar{3}$m}$\mathtt{\rightarrow}$\emph{Pnma} transition pressure is 26.5 GPa, according well
with recent experimental results of $\mathtt{\sim}$30 GPa. Our calculated transition-accompanied volume collapse
of 5.9\% is also in good agreement with the experimental data of 6.1\%. Under more higher pressure, we further found
that there occurred an isostructural transition between 80 and 130 GPa for the \emph{Pnma} phase, which is to be experimentally verified in future.

\begin{acknowledgments}
This work was supported by the Grant Agency of the Chinese Academy
of Engineering Physics. \end{acknowledgments}

\end{document}